
\magnification 1100


 \def\za{\alpha} \def\zb{\beta} \def\zc{\gamma} \def\zd{\delta}
 \def\ze{\varepsilon}   
   \def\zn{\nu}

 \def\zC{\Gamma} \def\zD{\Delta} \def\zF{\Phi} 
 \def\zO{\Omega}


     \def\cC{{\cal C}}  
       
     \def\cK{{\cal K}}  \def\cL{{\cal L}}
 \def\cM{{\cal M}}  \def\cN{{\cal N}}    \def\cP{{\cal P}}

 \def\cX{{\cal X}}  


   \def\dC{I\!\!\!\!C}    
       \def\dI{I\!\!I}
 \def\dJ{J\!\!\!J}      
 \def\dN{I\!\!N}      
     \def\dZ{Z\!\!\!Z}


 \def\dd{\partial}


 \def\dx{\triangle(X^-_1)}
 
 \def\cLv{\cL^{(h)}}

 \def\fy{f_0} \def\fz{f_1}

\vfill\eject

\hfill                                             SISSA -- 111/92/EP
\smallskip
\hfill                                             hepth@xxx/yymmnn

\vskip 3cm
\centerline{\bf    REDUCTION OF THE  KNIZHNIK  - ZAMOLODCHIKOV EQUATION}

\centerline{\bf  -- A WAY OF PRODUCING VIRASORO SINGULAR VECTORS
\footnote{${}^*$}{\rm Work  supported in  part by the Bulgarian Foundation for
Fundamental Researches
 under contract $\phi -11 - 91. $}}
\vskip 2cm
\centerline{\bf A.Ch. Ganchev\footnote{$^\dagger$}
{\rm postoctoral fellow at 1); mail address - 3); permanent address - 2)}
 and V.B. Petkova\footnote{$^{\dagger\dagger}$}
{\rm visitor at 1) and 3); permanent address and address after May 1992 - 2)}}
\bigskip
\centerline{1) Istituto Nazionale di Fisica Nucleare, Sezione di Trieste}
\smallskip
\centerline{2) Institute for Nuclear Research and Nuclear Energy,
Sofia 1784, Bulgaria}
\medskip
\centerline{3) SISSA-ISAS, via Beirut 2-4, 34014 Trieste, Italy}
\vskip 2cm

\centerline{\bf Abstract}
\vskip 1cm

We prove that for (half-) integer isospins the $sl(2,C\!\!\!\!I)$
Knizhnik - Zamolodchikov equation reduces to the
decoupling equation coming from Benoit - Saint-Aubin singular vectors.
In the general case an algorithm is suggested
which transforms, via the Knizhnik - Zamolodchikov equation, a Kac - Moody
singular vector into a Virasoro one.

\footline={\hfill}
\vfill\eject
\pageno=1\footline{\hfil\folio\hfil}

\noindent{\bf 1.\ }
Recently  an algorithm  was proposed
 [1]  for obtaining the singular vectors
of the  reducible Virasoro (Vir) algebra Verma modules
$V_{c,h}$ parametrised by

$$
  c(k)=13 - 6(k+2 + {1\over k+2})\,,
  \quad   h(J)=h_{r,r'}=J(J+1)/(k+2)  -J\,,
\eqno(1.1)$$
$$
   J=j-j'(k+2)\,, \qquad r=2j+1, \ r'=2j'+1 \,.
$$
with  $\, 2j, 2j' \in \dZ_+ \,$ and $k \in C\!\!\!\!I$, $k \neq -2$.
Rather involved in general, the algorithm simplifies in the particular
subseries
with $r=1$ (or $r'=1$ ) solved earlier [2]. In the  alternative
reformulation of this result in [1],
inspired by the Drinfeld - Sokolov formalism [3],
the highest weight  (vacuum) state appears embedded in an auxiliary
 $sl(2)$ -  multiplet. The singular vector is
  recovered recursively from a
matrix equality linearly depending on the Virasoro generators.

In an independent study [4]  of the quantum reduction of the
conformal WZNW models [5], based on the algebra $A_1^{(1)}$ at
arbitrary level $k$
 and isospin values $J$ as in (1.1),
 the general $n $- point chiral correlators
 were constructed
as solutions of the Knizhnik - Zamolodchikov (KZ) system of
equations [6], generalizing earlier results [7], [8].
The appropriate basis used in [4] to
describe the WZNW correlators
 allows to recover the corresponding
Dotsenko - Fateev (DF)  correlators  [9]
 of the reduced (Virasoro) theory with central charge and scale
dimensions    as in (1.1).   In the ``thermal''
case, described by integer or half-integer values  of the isospin
$J=j$,
this basis also allows the reduction of the
KZ system   to a higher order differential equation -- the BPZ equation [10],
accounting for the decoupling of the corresponding Virasoro
algebra singular vectors. It was argued in [4]
  that the analogy
with the matrix system in [1] is not accidental.

The aim of this paper is  first to
 prove that in the thermal case  the KZ
system can be recast in a matrix form that is equivalent, up to an explicit
``gauge transformation'', to the matrix form of the  singular vector given
in [1]. Next, developing further
 the argumentation of [4b], we show explicitly
 that the (infinite  in general)  KZ
matrix equation can be truncated, and hence reduced
to a higher order differential equation,  by imposing the algebraic
equation, originating from a  singular
vector of the Kac - Moody (KM) Verma modules. This allows
 us  to formulate -- for the time being as a
  conjecture --  an
  algorithm, checked on many examples, which transforms via the KZ
equation the KM singular vectors, found in
[11], into the general Virasoro singular vectors.

\medskip
\noindent{\bf 2.\ }
Let us denote by ${\cal A}$ the semidirect sum of the
 $A^{(1)}_1$ KM and the  Virasoro algebra with Sugawara central
charge. Our convention for the  $A^{(1)}_1$
 commutation relations is
$$
  [X^{\za}_n,X^{\zb}_m] = f^{\za\zb}_{\zc}\, X^{\zc}_{n+m} +
  k\, q^{\za\zb}\, n\zd_{n+m,0}\,, \quad
    f^{0\pm}_{\pm}=\pm 2\,, \, f^{+-}_0=1, \,
  q^{00}=2=2q^{+-}=2q^{-+}\,.
\eqno (2.1) $$

Consider a representation of $sl(2,\dC)\oplus sl(2,\dC)$ in a space of
functions $\cC_J$ of two complex
variables $x,z$ realized by the differential
operators
$$
  S^- = -{\dd\over\dd x}, \quad   S^0 = 2x{\dd\over\dd x} - 2J, \quad
  S^+ = x^2{\dd\over\dd x} - 2xJ\,,
\eqno (2.2)$$
and the generators  $-L_{-1},L_0,L_1$ of the
finite dimensional subalgebra of Vir given by the above
with the Sugawara dimension
$\zD_J=J(J+1)/(k+2)$ replacing the isospin $(-J)$
and the ``space'' variable $z$ replacing the ``isospin'' variable $x$.
Furthermore  $\,X^{\za}_n =z^n\,S^{\za}\,$ and  the standard differential
operators with respect to $z$ represent ${\cal A} $ with trivial centers.

To describe the solutions in [4] let us introduce an infinite set of
functions $W_t(x,z;J)$ with $t=0,1,2,\dots$
 and $x,z\in\dC^{n-1}$.
Set  $S=\sum_{a=1}^{n-1}J_a-J_n$ and
 denote by $\triangle(X)=\sum_{a=1}^{n-1}X_a$
the action of the generator $X$ in the tensor product of $n-1$ spaces;
$X_a$ is the action of $X$ at the $a$-th place and identity everywhere
else.

The functions $W_t(x,z)$ are subject to the conditions:
$$
  \triangle(S^-)\,W_t = 0 = \triangle(L_{-1})\,W_t \,,
$$
$$ \triangle(S^0)\,W_t= -2(J_n+S-t)\, W_t          \,,
\eqno (2.3) $$
$$
  \triangle(L_0)\, W_t = (\zD_n + S-t)\, W_t       \,,
$$
and the recursion relation
$$
  \dx\, W_{t+1} \equiv  \sum_{a=1}^{n-1}z_a  S^-_a\, W_{t+1}
   = \zn (S-t) (c_0-t)\, W_t    \,,
  \qquad     \dx\, W_0 = 0                        \,,
\eqno (2.4)$$
where $\, \zn =(k+2)^{-1}\,$ and $\,
   c_0 = \sum_{a=1}^n J_a + 1 - \zn^{-1}\,.$

They satisfy furthermore the system of  equations
$$
  K_{-1,_a} W_t(x,z) \equiv
  \left( {\dd\over\dd z_a} - \zn  \sum_{b(\ne a)}^{n-1} {\zO_{ab} \over z_{ab}}
  \right) W_t(x,z) =  {\dd\over\dd x_a} W_{t+1}(x,z) \,,
\eqno (2.5)$$

$$
  \qquad a=1,2,\dots,n-1 \,, \quad
    \zO_{ab} \equiv q_{\za\zb} S^{\za}_a S^{\zb}_b \,.
$$

Then the (infinite unless  $S$ is  a positive integer)  series
$$
  W(x,z) = const\sum_{t=0}^{\infty} ({-1 \over \zn})^t \,
{\Gamma(c_0-t+1) \over \Gamma(c_0-S+1)}\, W_t(x-z,z)\,,
\eqno (2.6)$$
represents the WZNW $n$-point primary fields
 correlators ``at infinity'', i.e., the functions
$$
  \lim_{x_n,z_n\to\infty} x_n^{-2J_n} z_n^{2\zD_n}
  \langle \zF^{J_n}(x_n,z_n)\dots\zF^{J_1}(x_1,z_1)\rangle
  =  \langle J_n,\zD_n |
  \zF^{J_{n-1}}(x_{n-1},z_{n-1})\dots\zF^{J_1}(x_1,z_1)|0 \rangle
$$
which determine the full $n$-point correlators up to a standard prefactor
if furthermore two of the points, say $(x_1,z_1)$ and $(x_{n-1},z_{n-1})$
are set to be $(0,0)$ and $(1,1)$ respectively.

For any set $J=(J_1,\dots,J_n)$ of isospin values such that
 $S=s-s'(k+2)$, $\,
s^{(')}=j_1^{(')}+j_2^{(')}+...+j_{n-1}^{(')}-j_{n}^{(')}\,,$
$s,s'\in\dZ_+\,,$
 a solution of (2.3)-(2.5) exists of the type
$$
  W_t(x,z) = \sum_{|\za|=t} \left(\prod_a{x^{\za_a}_a \over \za_a!}\right)
  \, \dI_{\za,\zC}^S(z)\,,
\eqno (2.7)$$
where $\za=(\za_1,\dots,\za_{n-1})\in \dZ^{n-1}_+$,
$|\za|\equiv \sum_{a=1}^{n-1}\za_a$
and $\dI_{\za,\zC}^S(z)$ are multiple $(s+s')$  integrals over a
cycle $\zC$ (see [4]
for the explicit expressions).
In particular $\dI_{0, \zC}^S=W_0\,$
coincide exactly with the correlators (at $z_n\!=\!\infty$) of
the Virasoro models with central charge $c(k)$ and
scale dimensions $\{h(J_a)\}\,$ given by (1.1).
The dependence on $\zC$ and $S$ will be omitted in what
follows.

 The integrals  $\dI_{\za}$
 satisfy a set of relations and a system of differential
equations which can be derived in a straightforward fashion
substituting (2.7) in (2.3-5). They generalize the
relations in [13], see also [8].
 The system of equations (2.5)  is equivalent to the KZ equation for the
$n$-point function $W(x,z)$. The solutions (2.7)
described by the integrals  $\dI_{\za}$
 generalize the solutions  [7], [8] of the KZ equations in
the standard WZNW theory corresponding to $s'=0\,,
S=s\in\dZ_+$; in that case   --
to be referred to as the
``thermal case''-- the sum in (2.6) is finite and
$W(x,z)=W_s(x,z)$.  The
integrals in (2.7)  can be
seen as  meromorphic modifications of the
general (``two - types screening
charges'') integrals of Dotsenko and Fateev [9] in the
 Virasoro theory described by (1.1).

The $n$-point correlation  functions
are invariant with respect  to the $sl(2,\dC)\oplus sl(2,\dC)$ subalgebra.
The corresponding Ward identities  carry over to
relations for $W$. The latter  read as
(2.3) if we set $t=S$ and formally identify $W_S$ with $W$
(although this identification is not necessary to prove the
relations for $W$).
The validity of these relations
 is ensured by (2.3), (2.4)
(or vice versa, together with (2.4) they imply (2.3)).
\footnote{${}^*$}
{The conditions (2.3) can be looked as selecting  l.w. states in the
tensor product $\cC_{J_1}\otimes\cC_{J_2}\otimes\dots\otimes\cC_{J_{n-1}}$,
i.e.,  $W_t$ are the singular vectors
generating, through the action of $\triangle(S^+)$,
 l.w. representations in this tensor  product.
The representations in  $\cC_{J_a}$, defined by the ``left'' action
generators (2.2),  are not
 necessarily  lowest  or highest weight   representations.}

The recursion relation originates in the Ward identity
corresponding  to $X^-_1$
and the fact that the state $(X^+_{-1})^l |J_n,\zD_n\rangle$ is a h.w. state
of $sl(2,\dC)\oplus sl(2,\dC)$
with highest weights $2(J_n+l)$ and $\zD_n+l$. This relation can be also
recovered combining the KZ system (2.5) with the
Ward identity due to $L_0$.
Similarly (2.5) ``intertwines'' the Ward identities related to
$S^-_0$ and $L_{-1}$ as is easily seen using the symmetry of
$\zO_{ab}$.

The recursion relation plays a crucial role. Indeed
already  in the standard case $s'=0$  it converts the conventional
 basis described by $\dI_{\za}$,
$|\za|=s$ (which originates in the Wakimoto bosonization technique)
  into one
 with an access to the ``reduced'' Vir theory,
i.e., into a basis containing $\dI_0$.
 The integrals are normalized in such a
way that in the thermal case they vanish unless $0\le|\za|\le
s_0$, $s_0={\rm min}(s, 2j_1, 2j_2,..., 2j_n)$.
According to (2.4) the operator $\dx$ acts as a raising operator in the
finite set of functions $\{W_t(x,z),\ t=0,1,\dots,s_0\}$.
As it is clear from (2.5) the role of the decreasing operator is played
by the KZ-operator in the l.h.s. of (2.5). More precisely, multiplying both
sides of (2.5) with $x_a$ and summing over $a=1,\dots,n-1$ we
obtain using (2.7)
$$
  (x\cdot K_{-1}) W_t = (t+1) W_{t+1} ,
  \qquad (x\cdot K_{-1}) W_{s_0} = 0 \,.
\eqno (2.8) $$
The commutator of the two differential operators $\dx$ and
 $(x\cdot K_{-1})$
reduces to a constant  on any $W_t$. Thus the finite set
$\{W_t\}_{t=0}^{s_0}$ can be viewed as a $sl(2)$ multiplet,
the lowest and highest states of
which represent the Vir and WZNW correlators $W_0$ and $W_{s_0}$,
respectively. In particular
$$
  (x\cdot K_{-1})^{s_0+1} W_0 = s_0! (x\cdot K_{-1}) W_{s_0} = 0
\eqno (2.9) $$
and this is the most compact form for the BPZ equation in the thermal
case. Indeed take for simplicity $s_0=2J_a=2j_a$ for
a given $a$. Expanding in powers of $x$ and selecting all terms at the
$2j_a+1$-th power of $x_a$  in (2.9)
one gets a $2j_a+1$ - order
partial differential equation, corresponding to the  singular vector
of the Virasoro Verma module labelled by the h.w. $h_{1,2j_a+1}$.

In the general (nonthermal) case the action (2.4) and (2.8) of the operators
$\dx$ and $(x\cdot K_{-1})$ is well defined, however,
there arise two different
infinite multiplets, one generated by  $(x\cdot K_{-1})$
starting from $W_0$ and
another -- by $\dx$, starting from $W$.

\medskip
\noindent{\bf 3.\ }
To recover the BPZ equations
 in a more transparent algebraic way let us introduce the
``right'' action of the algebra ${\cal A}$ in which any primary field
$\zF^{J_a}(x_a,z_a)$ is treated as a vacuum -- the h.w. state of a module
of descendants  [12], [10]. Namely, the subalgebra ${\cal A}_-\,$
spanned by $\, \{X_0^-, X_{-n}^{\alpha}, L_{-n}, n>0\}\,$
can be realized by the operators  $\cX^-_{0,a}= - S^-_a$ and
$$
  \cX^-_{-n,a}=\sum_{b(\ne a)} { S^-_b \over z_{ba}^n}\,,
  \qquad
  \cX^0_{-n,a}=\sum_{b(\ne a)} { S^0_b+2x_a S^-_b \over z_{ba}^n}\,,
  \qquad
  \cX^+_{-n,a}=\sum_{b(\ne a)}
  { S^+_b-x_a S^0_b-x_a^2 S^-_b \over z_{ba}^n}\,,
\eqno (3.1) $$
$$
  \cL_{-n,a} =  \sum_{b(\ne a)} {1 \over z_{ba}^{n-1}}
      \left(  {(n-1)\zD_b \over z_{ba}}
-{ \dd\over\dd z_b} \right),
  \qquad     \zD_b = \zn J_b(J_b+1)\,.
\eqno (3.2) $$
In terms of the generators (3.1) the KZ equations (2.5) read
$$
  \cK_{-1,a} W_t \equiv
  \left( \cL_{-1,a} - \nu (\cX^+_{-1,a}\cX^-_{0,a}+J_a\cX^0_{-1,a}) \right)
  W_t = \cX^-_{0,a} W_{t+1}\,,
\eqno (3.3)$$
where $ \cK_{-1} = K_{-1} $.
We shall often omit the label $a$. Next consider
$$
  \cK_{-n,a} =  \sum_{b(\ne a)}^{n-1} {(-1)^n \over z_{ab}^{n-1}} \cK_{-1,b}
  \,, \quad n>1\,,
\eqno (3.4) $$
and denote by $\cM_{-n}$ the finite-sum piece of the Sugawara formula for
the Vir generators that survives when acting on a h.w. state, i.e.,
$$
  \cM_{-n,a} \equiv {\zn\over 2} \sum_{k=1}^{n-1} q_{\za\zb}
  \cX^{\za}_{-n+k,a} \cX^{\zb}_{-k,a}
  + \zn ( \cX^+_{-n,a} \cX^-_{0,a} + J_a \cX^0_{-n,a} )\,.
\eqno (3.5) $$
One easily obtains that $\cK_{-n}$ defined in (3.4) can be rewritten as
$$
  \cK_{-n,a} = \cL_{-n,a} - \cM_{-n,a}\,.
\eqno (3.6) $$
Indeed, using that $\zO_{ab}$ is symmetric we have
$$
   \sum_{b\ne c(\ne a)} {2\zO_{bc} \over z_{ab}^{n-1}z_{bc}} =
   \sum_{b\ne c(\ne a)} {\zO_{bc}\over z_{bc}} \left( {1\over z_{ab}^{n-1}}
   - {1\over z_{ac}^{n-1}} \right) =
   \sum_{k=1}^{n-1} \sum_{b,c(\ne a)}
   {\zO_{bc} \over z_{ab}^{n-k}z_{ac}^k}
   - (n-1) \sum_{b(\ne a)}  {\zO_{bb} \over z_{ab}^{n}}\,,
$$
which gives (3.6) when substituted in (3.4). Inserting
 (3.3) into (3.4)  and using (3.1) we obtain
$$
  \cK_{-n,a}\, W_t = \cX^-_{-n+1,a}\, W_{t+1} \,,
\eqno (3.7) $$
a form of the KZ equation which will be the most useful in what follows.
Written in terms of $\,\cL_{-n}\,$
the equality (3.7) is the analogue of the Sugawara formula
for the negative grade subalgebra of Vir. In terms of the set
$\{W_t\}$ this formula acquires
an additional term in the right hand side. As we shall see this
allows
to effectively ``invert'' the Sugawara construction, i.e., to convert the
affine KM generators $\cX^-_{-n}$
 into Virasoro ones.

The set of equations (2.5) (or (3.3)) can be reproduced starting from the
standard form of the KZ equation
$$\,  K_{-1}\, W(x,z) = 0\,,
\eqno(3.8)$$
with $\,W\,$ defined by the $x-z$ expansion (2.7) and using the linear
relations (2.3a), (2.4) [4].
On the other hand one can write down a straightforward simple, although
formal equation, encompassing (2.5). Namely, let us identify $W$ with
$W_S$ by some way of analytic continuation giving meaning to
``nonmeromorphic'' integrals $\dI_{\za}\,,$ with some $\, \alpha_b
\not \in \dZ\, $ (such possibilities were discussed
in [14] and [4]). Then one can  apply the formal $S-t$-th power
of the intertwining operator $\,\dx\,,$ so that according to (2.4)
we can write
$$
  \dx^{S-t} K_{-1}\, W_S(x,z) = 0 \,.
\eqno (3.9) $$
Although both $W_S$ and $\dx^{S-t}$ are formal the resulting expression is
well defined and reproduces (2.5).
As will become clear this trick will be extremely useful in treating also
the algebraic equations due to KM singular vectors in the general
nonthermal case.

Similarly, multiplying the basic system of equations (2.5)
with $z_a^n$ and summing over $a=1,2,...,n-1,$ one gets an
expression for $\triangle(L_n), n\geq 0$ analogous to (3.7), (3.6) with
$\cX_{-p+1}^{-}$ replaced by $\triangle(X_{p+1}^{-})$ and
$\cX_{-p}^{\alpha}$ in $\,   \cM_{-m,a}\,,$ (3.5),
 replaced everywhere by $\triangle(X_p^{\alpha})$.
\medskip
\noindent{\bf 4.\ }
 Our next objective is to obtain from the above a matrix equation
for the integrals $\dI_t\equiv \dI_{t\ze_a}$ for a fixed $a$
(here $(\ze_a)_b=\zd_{ab}$).
Obviously $(\cX^-_{0,a})^t\,W_t=\dI_t$. Thus
applying $(\cX^-_{0,a})^t$ to both sides of (3.7) and
commuting it through $\cK_{-n}$
we obtain
$$
   (\cX^-_{0,a})^t  \cX^-_{-n+1,a} \, W_{t+1} =
  (\cK_{-n,a} + \zn t \cX^0_{-n,a}) \dI_t
  - \zn \zc_t(J_a)\, (\cX^-_{0,a})^{t-1} \cX^-_{-n,a} \, W_t
\eqno (4.1) $$
where $\zc_t(J)=t(2J-t+1)$.
Eliminating recursively the terms
$\, (\cX^-_{0,a})^t  \cX^-_{-n+1,a} \, W_{t+1} \,$ we get
$$
  \dI_{t+1} = (\cK_{-1}+\nu t \cX^0_{-1}) \dI_t +
  \sum_{p=1}^t (-\nu)^p \prod_{i=0}^{p-1} \zc_{t-i}(J)\,
  (\cK_{-p-1} + \nu (t-p)\, \cX^0_{-p-1}) \, \dI_{t-p}\,.
\eqno (4.2) $$
The above equation can be alternatively  obtained by using directly the KZ
equation written  in terms of the integrals $\dI_{\za}$.

The relation (4.2) is valid for any $t=0,1,2\dots$ and obviously can
be cast in a triangular (in general infinite) matrix form.
Introduce generators $\dJ^{0,\pm}$ with the same commutation relations
as $S^{0,\pm}$ acting on a basis $\, \{v_t, t=0,1,...\}\,$ as
$\ \dJ^-v_t=v_{t+1}\,, $
 $\ \dJ^+v_t=\zc_t(j)v_{t-1}$, $\ \dJ^0v_t=2(j-t)v_t\,, $
and set
$$
  \dI(v,z) = \sum_{t=0} v_{2j -t} \dI_t(z)\,.
\eqno (4.3) $$

 Let
us now restrict ourselves to the case $2J_a=2j_a \in \dZ_+$ in which
$\dJ^{0,\pm}$ can be realised by finite matrices (assuming
 $\dJ^- v_{2j}=0\,$)  and
 the set $\{\dI_t\}_{t=0}^{2j_a}$ can be
turned into a $2j_a+1$-dimensional $sl(2,\dC)$ module.
In this basis (4.2) can be written as
$$
  \cK \dI \equiv  \left( -\dJ^- + \sum_{p=0}^{2j} (-\nu \dJ^+)^{p} \,
  (\cK_{-p-1} + \nu{\cX^0_{-p-1}\over 2}
   (\dJ^0 + 2j)) \right) \dI = v_0 \dI_{2j+1}\,.
\eqno (4.4) $$
We keep the r.h.s.  although it is identically zero;
the same formula with summation not restricted from
above and with zero r.h.s. holds for arbitrary $J$.

Following [1] let us introduce the matrix system
$$
  \cL F \equiv \left( - \dJ^- +
  \sum_{p=0}^{2j} (-\nu \dJ^+)^{p} L_{-p-1}
   \right) F =v_0 F_{2j+1}\,,
\quad   F= \sum_{t=0}^{2j} v_{2j-t} F_t \,,
\eqno (4.5) $$
or, in components
$$
F_{t+1}= \sum_{p=0}^t (-\nu)^p \prod_{i=0}^{p-1} \zc_{t-i}(j)\,
  L_{-p-1}\, F_{t-p} = \cN_{t+1}(j) F_0\,,
\eqno(4.6)$$
where $\,\cN_{t}(j)\, $ is  a polynomial in the (negative grade)
 Vir generators,
 $$
  \cN_t(j) = \prod_{i=1}^{t-1} \zc_i(j) \sum_{r=1}^{t} (-\nu)^{t-r}
  \sum_{k_i\geq 0: \sum_{i=1}^r k_i = t-r}
  { L_{-1-k_r} \dots L_{-1-k_1} \over
  \zc_{k_1+\dots+k_{r-1}+r-1}(j) \dots \zc_{k_1+k_2+2}(j)
  \zc_{k_1+1}(j) } \,.
$$

In particular if $F_0$ is the highest weight state of the
reducible Vir Verma module
$\,V_{ c(k)\, h(j)}\,,$ $\, 2j  \in \dZ_+\, $ (i.e.,
$L_0 F_0= h_{2j+1,1} F_0\,,$ $L_n F_0 =0 \,$ for $  n>0 $) then
 $F_{2j+1}$ reproduces [1] the expression found in [2]
  for the singular vector of weight $h(j)+2j+1$.
 We can realize the abstract Vir generators $L_{-n}\,,
n>0$  by the differential operators $\cLv_{-n,a}$
defined as in  (3.2) with the Sugawara conformal
dimensions $\zD_{J_a}$ replaced by $h(J_a)$.
Clearly the system (4.4) ( (4.2) ) derived from the KZ system of equations
 has almost the same structure as
(4.5) ( (4.6) ).
 Indeed on $\dI_t$, which are independent of $x$, the operator
$\cK_{-n}$ reduces to
$$
  \cK_{-n} \dI_t =
  \left[ \cLv_{-n} - {n-1\over 2} \cX^0_{-n} -
  {\nu \over 4} \left( \sum_{k=1}^{n-1}
   \cX^0_{-n+k}\cX^0_{-k} - 2(n-1-2J) \cX^0_{-n}
  \right)  \right] \dI_t \,,
\eqno (4.7) $$
and furthermore $\cX^0_{-n}$
reduces to  $\hat {\cX}^0_{-n}\equiv \sum_{b(\ne a)} {-2J_b\over z_{ba}^n}
\,$,  $\,\hat {\cX}^0_{-n} \dI_t = \cX^0_{-n} \dI_t$.
This suggests that the  functions
 $\{F_t\}$ can be realized as $z$-dependent linear combinations of the
 integrals  $\{ \dI_t \}$. Indeed define
 $$
  F=g\dI\,,
\eqno (4.8) $$
where
$$
  g_a =\left[ \exp(-\nu \dJ^+ (L_{-1,a}-{1\over 2}\cX^0_{-1,a}) )
  \right] \cdot 1
\eqno (4.9)$$
$$
=1\!\!1+\nu \dJ^+ {\hat {\cX}_{-1}^0 \over 2} + { (\nu \dJ^+)^2 \over 2}
\left[\left({\hat {\cX}_{-1}^0 \over 2}\right)^2
- {\hat {\cX}_{-2}^0 \over 2}\right]\, +\, \dots
$$
and the sum is finite  for $r=2j+1$ - positive integer, since
 $\, (\dJ^+)^{2j+1}=0$.
\smallskip

{\bf Proposition 1.} {\it The  matrix systems (4.4) and (4.5)
 are connected by the "gauge transformation" $g$, i.e.,}
 $$
  g \cK\, \dI = \cL g\, \dI\,  \,.
\eqno (4.10) $$
The proof consists of the following two Lemmas.

{\bf Lemma 1.}
$$
  g \left( - \dJ^- + \sum_{k=1} (-\nu \dJ^+)^{k-1} \,
  ( -\cM_{-k,a} + \nu{\cX^0_{-k,a}\over 2} (\dJ^0 + 2j_a)) \right) \dI
     = - \dJ^- \, g \, \dI \,.
\eqno(4.11)$$
Denote
$\,  R_{-n} \equiv \left({1\over 2}\cX^0_{-1,a}-L_{-1,a}\right)^n \cdot 1 \,$
and use the $ sl(2)$ commutation relations of the generators
$\dJ^{0,\pm}$ to get
$$
  [-\dJ^-,g ]   = \nu \sum_{k=1} (\nu \dJ^+)^{k-1}
 { R_{-k} \over (k-1)! }  \left( \dJ^0  +  k-1 \right) \,.
$$
The lemma is obtained  using twice the following identity   proved
 by induction
$$
   { R_{-k} \over (k-1)! }
  = \sum_{l=0}^{k-1}
   {(-1)^{l+k-1}\over l!} R_{-l} {  \hat {\cX}^0_{-k+l,a} \over 2} \,.
$$

{\bf Lemma 2.}
$$
  g(u) \left(\sum_{n=1}\cL_{-n,a}u^{n-1}\right) g(u)^{-1}
    = \sum_{n=1} \left( \cL_{-n,a} + (n-1) { \hat
{\cX}^0_{-n,a} \over 2}
 \right)u^{n-1} \,
  \equiv \sum_{n=1}\cLv_{-n,a}u^{n-1} \, .
\eqno(4.12)$$
Here $\, g(u)=[ \exp(u(L_{-1}-{1\over 2}\cX^0_{-1}) )
  ] \cdot 1 \,$. To prove the lemma one has to use
$$
  {\dd\over\dd z_b} R_{-n,a}  = \sum_{k=1}(-1)^{k+1} {n\choose k}(k-1)!
  R_{-n+k,a}
    {\dd\over\dd z_b} \,{ \hat {\cX}^0_{-k,a}\over 2}\,.
$$

In terms of the coefficient functions $\dI_t$ and $F_t$  (4.8) reads
$$
 \cN_t(J)\dI_0 \equiv   F_t =\dI_t + \sum_{p=1}^t {\nu^p\over p!}
  \left(\prod_{i=0}^{p-1} \zc_{t-i}(J) \right)
  \left( R_{-p} \right) \, \dI_{t-p}\,.
\eqno(4.13)$$
The gauge transformation (4.8) keeps invariant the first and the
last elements of the multiplet $\{F_t\}$, i.e., $F_0=\dI_0, F_{2j+1}=
\dI_{2j+1}$. For $t=2j+1$ each term in the sum in (4.13) vanishes since
$\zc_{2j+1}(j)=0\,$ so that $\dI_{2j+1}$ (which is identically
zero in our realization) can be identified with the singular
vector, given explicitly by the l.h.s., i.e., $ \cN_{2j+1}(j)\dI_0$.
The splitting (4.13)
of  any $\dI_t\,, t=0,1,...,\,$ as a sum of two terms,
 hold for  arbitrary values of $J$
- then  $\dI$ and $F$
become infinite series while
the r.h.s. of (4.4) and (4.5) should be replaced by zero.

\medskip\noindent{\bf 5.\ }
 For spins  of the form
$J=j-j'/\nu$ with $r=2j+1$, $r'=2j'+1 \in \dZ_+$, $\nu \in
C\!\!\!\!I$, $\nu \neq 0\,,$
 the KM Verma module of highest weight $2J$  contains a singular vector
of weight $\, 2(J-r)\,$ (this is part of the Kac-Kazhdan theorem [15]).
There is an explicit expression for the singular
vector $P_{(r,r')} |J\rangle$ given by the Malikov Feigin Fuchs (MFF) formula
[11]:
$$
  P_{(r,r')} = \fz^{r+(r'-1)/\nu}\, \fy^{r+(r'-2)/\nu}\, \dots \,
  \fz^{r-(r'-3)/\nu}\, \fy^{r-(r'-2)/\nu}\, \fz^{r-(r'-1)/\nu}\,,
\eqno (5.1) $$
where for short we denote  $\fz=X^-_0$, $\fy=X^+_{-1}$.
This expression is a monomial in $\fz,\fy$ but they are raised to,
in general,   complex powers. It can be viewed as a compact notation
for a polynomial in $X^-_0, X^{\za}_{-n}$, $n=1,2\dots$,
with coefficients that are polynomials in $1/\nu$.
In practice, to obtain such an expanded form one can start
from the middle of (5.1) where $\fy$ or $\fz$, depending on whether
$r'$ is even or odd respectively, is raised to the integer power $r$. With
the help of
$$ \fy^{2+q} X^-_{-m} \fy^{-q} =  \fy^2 X^-_{-m}
  + q \fy X^0_{-1-m} + q(q-1) X^+_{-2-m},
  \qquad    \fy^{1+q}  X^0_{-m} \fy^{-q} = \fy X^0_{-m}
   -2q X^+_{-m-1}
$$
$$
  \fz^q X^+_{-m} \fz^{2-q}  =  X^+_{-m} \fz^2
  -q X^0_{-m} \fz - q(q-1)  X^-_{-m},
  \qquad
   \fz^q X^0_{-m} \fz^{1-q} =
   X^0_{-m} \fz +2q X^-_{-m}
\eqno (5.2) $$
we can move from the middle of (5.1) outwards eliminating all noninteger
powers. For a different algorithm to obtain the singular vectors using
fusion see [16]. It is in principle
also possible to write (5.1) as a homogeneous polynomial
in the generators $\fz \,,\fy\,,$  of
degree $\, rr'\,,$ and $\,
 r(r'-1)\,,$ respectively (see [17] for such explicit expression
in the  case $\,(r,r')=(r,2)\,$).
Substituting in $P$ (to be denoted by $\cP$)
 the realization (3.1) of the generators $\,f_0,\, f_1 \,$
we impose the condition
$$
  \cP_{(r,r'),a} W (x,z) = 0\,,
\eqno(5.3) $$
which expresses the  decoupling of the corresponding
  descendant of the primary field
$\zF^{J_a}(x_a,z_a)$  --  we can extend (3.1) setting
$  \cX^+_{0,a} W = 0 $, $\,
  \cX^{\alpha}_{n,a} W = 0, $ for $n> 0 \,,
\alpha=0\pm\,, $ $\cX^0_{0,a} W = 2J_a  W\,.$
 \footnote{${}^{**}$}
{Alternatively, one can impose the algebraic condition on the
state $\langle J_n, \triangle_n |\,$ ; then the condition on $\,W\,$
looks like (5.3) but
  $\, \cX_0^-\,$ and $\,\cX_{-1}^+\,$ in  $\,\cP_{r,r'}\,$
 are replaced by $\,\triangle (S^+)\,$ and $\,   \triangle (X_1^-)\,$
respectively.}

The algebraic condition (5.3) is equivalent
to linear (with rational in $z$ coefficients) relations
for the set of functions $\{W_t(x,z)\}$, or, equivalently
for the integrals $\,\{\dI_{\mu}\}\,$. Namely for
 any $\, t\geq rr'\,  $ we have
$$
{(\cX_{0}^-)^{rr'} \over (r(r'-1))!} W_t\,
 +\,  \sum_{k=0}^{r(r'-1)-1} { \nu^{r(r'-1)-k} \over k! }
  \cP_{(r,r')}^{(k)} W_{t-r(r'-1)+k} = 0,
  \qquad
  P^{(k)} \equiv \left({\dd\over\dd \fy}\right)^k P \,.
\eqno (5.4) $$
Note that $P^{(k)}$ is well defined, i.e., does not depend on the
form chosen to represent $P$. Indeed, starting from $P$ written as
a polynomial only in $\fz,\fy$ we can obtain any other form by multiple
commutations. At each step we will use
$\fy X^-_{-n} = X^-_{-n} \fy + X^0_{-n-1}$, $n\ge 0$, or
$\fy X^0_{-m} = X^0_{-m} \fy - 2 X^+_{-m-1}$, $m\ge 1$, which remains an
equality after ``differentiating'' in $\fy$ (it is important that $m\ne 0$).
In particular, the highest derivative  reproduces the first
term in (5.4).
In the thermal case $\, r'=1\, $ this  is the only term
 and the equation
(5.3) is automatically satisfied since $\, W(x,z) \,$ is a
polynomial of highest degree $\,r-1=2j\,$ in $\,x\,$.

The easiest way to  obtain (5.4) is to start,
as in the derivation of the KZ equation (2.5)  for $W_t$,
 from the ``nonmeromorphic''
$W_S$ and then  ``intertwine'' it to a ``meromorphic'' one, i.e., consider
$$
  \dx^{S-q} \cP_{(r,r')} W_S = 0, \qquad
  q = t - r(r'-1) \geq r\,.
\eqno (5.5) $$
With the help of
$$
  [\dx^A,\fy^B] = \sum_{k=1}^{\infty}
  {(\dd_{\fy}^k\,\fy^B)\over k!}  \left( \prod_{j=0}^{k-1}
  (A-j)\,(2k-j-2J-A-B-\triangle(S^0)) \right) \, \dx^{A-k}
\eqno (5.6) $$
computed with $\fy$ being realized as $\cX_{-1}^+\,,$
we can move $\dx$ through $P_{(r,r')}$. For example, to obtain the
term with first ``derivative'' in (5.4) apply (5.6) to each factor
$\fy^{B_{i}}\,$ of (5.1), ($B_i\equiv r-(r'-2i)/\nu$, $i=1,\dots,r'-1)$,
keeping only the first term of (5.6). Commute
$(\triangle(S^0)+B_{i}+\dots)$
to the right through the powers of $\cX_{-1}^+\,,$ and $\cX_{0}^-\,,$
and apply at the end the second equality in  (2.3).
 The dependence on $i$ cancels and (5.5) becomes
$$
  0 = \cP \dx^{S-q} \, W_S +
 (S-q)(c_0-q)\,\cP' \, \dx^{S-q-1} W_S + \dots
$$
and it remains to use the recursion relation (2.4) to recover the
``first derivative'' term in  (5.4). The higher ``derivative''
 terms are obtained in a similar fashion.

 It remains an open problem to prove, using the explicit
expressions for the integrals $\dI_{\za}$ found in [4],  that  indeed the
relations (5.4), or equivalently, the condition on the correlators (5.3)  hold.
For the time being we are able to check this in the simplest
examples so the validity of (5.4) will be taken as an assumption.

The equality (5.4) implies that $\dI_t$
 for every $t\ge rr'\,$ can be expressed in terms of $\dI_{\za}$
with $\,r \le |\za|<t\,$.
The important relation   is the one for $t=rr'$, which
for $\, r'>1\,$ is transformed
into the corresponding BPZ equation as we will argue below.
 Plugging the expressions for the
right action into (5.4), (5.1) for $t=rr'$ the dependence on $x$
disappears; the first term reproduces
 $\dI_{rr'}$.
For example,  in the simplest nonthermal case
$(r,r')=(1,2)$, i.e., $2J_a=-1/\nu$, the linear relation is:
$$
 \dI_{2\ze_a}=  (1+\nu) (\cX_{-1}^0\, \cX_0^- -\gamma_1(J)\,
  \cX_{-1}^-)\, W_1  =  (1+\nu) \sum_{b(\ne a)}
  {2J_a \dI_{\ze_b} - 2J_b \dI_{\ze_a} \over z_{ba}}\,.
\eqno(5.7)$$
Another example is provided by
$(r,r') = (2,2)\,,$ i.e., $2J=1-1/\nu$:
$$
\dI_4 + \nu  \cP_{(2,2)}^{(1)} W_3+ {\nu^2 \over 2}
\cP_{(2,2)}^{(0)} W_2=0\,, \qquad
\cP_{(2,2)}^{(1)} W_3={\zc_4 \over 2}
[\cX_{-1}^0 \dI_3 -{\zc_3 \over 3}
(\cX_{0}^-)^2\, \cX_{-1}^-\, W_3]\,,
\eqno(5.8)$$
$$
\cP_{(2,2)}^{(0)} W_2  =
 { \zc_3\zc_4 \over 3!}\left[
    ((\cX_{-1}^0 )^2  +  \cX_{-2}^0 ) \dI_2
-{\zc_2} \left(  \cX_{-1}^0 \cX_{0}^-  \cX_{-1}^-
+  \cX_0^- \cX_{-2}^-
-{\zc_1 \over 2} ( \cX_{-1}^-)^2 \right) W_2  \right].
$$
Recalling  the realization of $\,\cX_{-n}^0\,
\,, \,\cX_{-n}^-\,$ in (3.1) we see
that (5.8) is a linear combination of $\, \dI_{\alpha}\,$ for
$\,2 \le |\alpha|\le 4\, $.

It is convenient to rewrite in general  the singular
vector in ``$+$$0$$-$''-ordered form with $X^+$ to the left,
$X^0$ in the middle and $X^-$ on the right. Then
just by counting the degree in $x$
 one sees that for $\, t=rr'\,$ in (5.4) all therms with $X^+$ give zero
 and we are left with terms of the type   ``$0$$-$'' as in (5.7),
(5.8).
Now
solving (5.4) for the first term $\, \dI_{rr'}\,$
and  inserting  into (4.2) ( or (4.13))  one
effectively
 truncates the infinite KZ system. Exploiting repeatedly
 (3.7)  for $\,t<rr'\,$
   (or the relations (4.1), (4.2), (4.13) derived as a  consequence of it)
    it can be furthermore reduced to higher ($rr'$ - )
order partial  differential equation for $\dI_0$ -- a BPZ - type equation,
accounting for the decoupling of the   singular vector
of the Virasoro module $V_{c(k)\, h(J)}$. With the
help of (3.7), (4.7) these equations  can be written  in terms of the
generators $\, \cLv_{-n}\,$ acting on the minimal model integral
$\dI_0$.
It seems natural to expect, although technically  difficult  to
prove, that the rest terms, similarly to
 the gauge degrees of freedom in Proposition 1,
cancel in the final equation for $\dI_0$. Thus we conjecture
the following

\medskip\noindent
{\bf Proposition 2. }
{\it For any $\, 2J+1=r - (r'-1) / \nu \,$,
$\,  \nu \neq 0, \infty$, and $r,r'-1\in\dN$,
the set of equations (5.4) for $\,t=rr'$, accounting for the
decoupling of the KM null vector, and (3.7) for
$t=0,1,..., rr' -1$, equivalent to the
KZ system,  determine the singular vector at level $rr'$ of the Virasoro
Verma module  $\,V_{c(k), h(J)}\,$.}
\smallskip

We recall that in the thermal case $\, r'=1\,$ the KZ system (3.7) itself
reduces to the corresponding Virasoro singular vector, while
the KM algebraic condition (5.4)  is trivially satisfied, i.e.,
in this case the two conditions decouple.

Let us illustrate the mechanism of reduction on the  examples
considered  above.
Using  the splitting (4.13) for $\dI_2$ and (3.7), (4.7)
to convert  $  \cX_{-1}^-\,$ we obtain that (5.7) is equivalent to
$$
 \cN_2(J)\,\dI_0 \equiv \left((\cL^{(h)}_{-1})^2
  - \nu \zc_1(J)\, \cL^{(h)}_{-2}\right)\,\dI_0 =
-(1+\nu) \zc_1(J)\, \cL^{(h)}_{-2}\,\dI_0\,,
\eqno(5.9)$$
Indeed, apart from  the term proportional to $\cLv_{-2}\,\dI_0\,$,
  the r.h.s of (5.7) exactly compensates
the term containing gauge degrees of freedom  in  the r.h.s. of (4.13).
Taking into account that  $\zc_1(J) =2J=-1/\nu$
 (5.9) recovers the corresponding Vir singular vector.

In the next example $(r,r')=(2,2)\,,$ i.e., $2J=1-1/\nu \,,$
let us neglect for the time being all terms in (5.8),  which in
the "$0$$-$" - ordered form  start with some $X_{-n}^0$, i.e.,
$$
  \dI_4 -
  {\nu\over 3!} \zc_4\zc_3 (\cX_0^-)^2 \cX^-_{-1}\, W_3
  - { \nu^2\over 4!} \zc_4\zc_3\zc_2 \left( 2 \cX_0^- \cX^-_{-2} \, W_2
- \zc_1  (\cX^-_{-1})^2)\right) \, W_2
\approx 0\,.
\eqno (5.10)$$

We now  apply the KZ equation in the form (3.7) repeatedly
untill we turn all $X^-_{-n}$ into $K_{-n-1}$. For the terms of the
form $(\cX_0^-)^n \cX^-_{-k}W_{n+1}$ we can use directly (4.1). For the
term involving $\ (\cX_{-1}^-)^2\,$ we need  the commutator
$$
  [X^-_{-1},K_{-2}] = -(\nu 2J+1) X^-_{-3}
  -\nu(X^0_{-3} X^-_0 + X^0_{-2} X^-_{-1} - X^0_{-1} X^-_{-2} ) \,,
$$
so that $\, (\cX^-_{-1})^2 W_2 \approx ( \cK_{-2}^2 - \nu
\cK_{-4}) \dI_0 \,$ up to the neglected terms.
   Using (4.7) keep for
  any $\, \cK_{-n}\,$  in the l.h.s. of (5.10)
only the term $\,\cLv_{-n} \,$ while  for
the integrals $\,\dI_t\,$ one has to use
the splitting (4.13) putting aside the second term in the r.h.s
of (4.13).
Finally a tedious computation shows that all the  neglected terms
 combine and give  zero, so that we get
$$
  \left[ \cL^{(h)\,4}_{-1} + (\nu-{1\over\nu})^2\, \cL^{(h)\,2}_{-2}
  - 2(\nu+{1\over\nu}) \cL^{(h)}_{-2} \cL^{(h)\,2}_{-1}
  +2(3-\nu-{1\over\nu}) \cL^{(h)}_{-3} \cL^{(h)}_{-1}
  - 3\nu(1-{1\over\nu})^2  \cL^{(h)}_{-4} \right] \dI_0 = 0 \,,
\eqno (5.11) $$
The  l.h.s. of (5.11) recovers the  expression for  the Virasoro  singular
vector at level 4 of the Virasoro Verma module $V_{ c(k)\, h_{2,2}}$.

The procedure illustrated  on these examples is in fact an  algorithm,
which can be implemented into a computer code
(e.g., we have used REDUCE). As generators of the subalgebra
$\,{\cal A}_-\,$ we choose  $\, X_0^-, X_{-n}^{\alpha}$ and the
``improved'' Vir generators
 $L^{(h)}_{-n}\equiv L_{-n}+{(n-1) \over 2}X_{-n}^0\,$ --
which reduce to $\cL^{(h)}_{-n}$ on $x$ - independent functions
in the realization for $L_{-n}$ and $X^{\za}_{-n}$ given by (3.2), (3.1).
Let us say that a monomial in these generators is ordered if it
has the form $X^+...X^0...L^{(h)}...X^-...\,$. The instructions we
have to provide are: 1) using the commutation relations order any
expression, 2) use (3.7), i.e.,  substitute
 $\, X_{-n+1}^- W_{t+1}\,$ by
$L_{-n}^{(h)}-{(n-1) \over 2} X_{-n}^0/2 -M_{-n}$ and $X_{-n+1}^-W_0$ by $0$.
Taking a particular KM singular vector we feed (5.4) into the
program and let it apply the instructions. It can be easily seen
that the  ordered terms containing $X^+$ actually do not survive.
One can speed up the algorithm
using along with (3.7)  its consequences -- the relations (4.1),
(4.2) or (4.13).
 We    have checked the above algorithm on several  examples.
 The result is a polynomial in the $L^{(h),}s\,$ acting on $W_0$
-- all ``gauge'' terms , i.e., containing $X^0$,  cancel.
Extending the negative grade subalgebra by setting
$\cL^{(h)}_{0,a}\,W_0=h(J_a)\,W_0$, $\,\,\cL^{(h)}_n\,W_0=0\,$,
for $\,n>0$, and
assuming the standard Vir commutation relations with central charge $c(k)$
it is straight forward to check that $\cL^{(h)}_1$, $\cL^{(h)}_2$
annihilate the final result.

  The proof of  Proposition 2 could be simplified if we were
able to rewrite (5.4) for $\, t=rr'\, $ in a matrix form which
can be combined with the matrix form of the KZ equation.
As in the proof of the  Proposition 1 above this would allow to
handle the gauge degrees of freedom and probably to obtain more
explicit expressions for the Virasoro singular vectors.
The matrix representation of [16] might be of use if rewritten in
the basis containing $\, \dI_0\,$ exploited here.

 There is
one special case in which such a matrix form appears naturally as a
consequence of the duality discussed in [4]. Namely, in the
``quasithermal'' case, described by $\, r=1\,\,$ (i.e., $\,J= -j'(k+2\,$)
one can use the dual WZNW theory -- defined in general by replacing
$\, J \to \hat J= -J/(k+2)\,, (k+2) \to ({\hat k} +2) = 1/ (k+2) \,$,
 which keeps invariant
 $\, ( c(k)\,, h(J) )\,$. In this case
the dual theory  is a thermal one, with $\hat J=j'$, and since the integral
$\,\dI_0\,$ is the same in both theories one can compare the
corresponding KZ systems. This produces a linear relation for
the quasithermal integrals $\, \dI_{\alpha}\,, |\alpha| \le 2j'+1\,$
which can be rewritten in a  matrix form
 in terms of
the two
sets of vectors  $\,\{\dI_t \,, t=0,1,.., 2j'+1\}\,$
and $\,\{{\hat \dI_t} \,, t=0,1,.., 2j'\}\,$,  using the matrix
  differential operators $   \cK\,, {\hat  \cK} \,$ (4.4)
    and the gauge transformations $\,  g\,, {\hat  g}\,$.

\medskip
\noindent{\bf 6.\ }
 We end with a few remarks.
 The marked difference
between the thermal ($r'=1\,$) and nonthermal cases
appears also in [5] where the thermal BRST
cohomology is trivial
\footnote{${}^{***}$}{We thank P. Bouwknegt for emphasizing this
fact to us.} and in [18] where going from  KM characters
one takes a limit in the thermal case and a residue in the  nonthermal one
to recover the Virasoro characters.

Clearly the mechanism of reduction   described here
   is essentially different from the standard procedure
discussed in [5] which requires fermions to implement
the action of the BRST operator on  Fock modules. Nevertheless, as
discussed in [4],
there is certainly some analogy in the role played by the
operator
  $\,\Delta(X_1^-)\,$
    and the standard BRST operator
used to impose the  constraint in [5].

Although the result in the general nonthermal case lacks so far the
rigour and the explicitness of the thermal one it looks
conceptually important that the problem of finding the
 singular vectors of the Virasoro Verma modules
is reduced to the analogous  simpler
problem for the KM algebra $A_1^{(1)}$. Generalizations seem
possible since
 the singular vectors for higher rank algebras are also known [11]
and  there are natural analogues of
the ``intertwining'' operator $\Delta (X_1^-)\,$ in the various
reduction schemes discussed in [5], [19].
In any case, the generalization of the result (Proposition 1)
 for the thermal subseries, i.e., dominant integral weights,
   should be easier.

Finally it would  be interesting to find a connection between the
algorithm discussed above and the
recently proposed  [20]
compact (though rather formal)
general formula for the Virasoro singular vectors,
inspired by the MFF expression.

\medskip

\noindent{\bf Acknowledgements}
\medskip

It is a pleasure to thank Paolo Furlan and Roman Paunov for
collaboration in the
reviewed work. The contribution   of
 R. Paunov, who   in particular
suggested  the compact formulae (2.8), (2.9), was essential for the
development presented here.

We  acknowledge the hospitality and the financial support of INFN,
Sezione di Trieste and SISSA, Trieste.

\medskip
\noindent{\bf References}
\medskip

\item{[1]}
   M. Bauer, Ph. Di Francesco, C. Itzykson and J.-B. Zuber,
    {\it Phys. Lett.} {\bf B260} (1991) 323;
    {\it Nucl. Phys.} {\bf B362} (1991) 515.
\item{[2]}
   L. Benoit and Y. Saint-Aubin, {\it Phys. Lett.} {\bf B215}
   (1988) 517.

\item{[3]}
   V. Drinfeld and V. Sokolov, {\it J. Sov. Math.} {\bf 30} (1984) 1975.

\item{[4]}
   P. Furlan, R. Paunov, A.Ch. Ganchev and V.B. Petkova,
   {\it Phys. Lett. } {\bf B267} (1991) 63; preprint CERN-TH.6289/91.

\item{[5]}
   M. Bershadsky and H. Ooguri, {\it Comm. Math. Phys.} {\bf
   126} (1989) 49;\par
   B. Feigin and E. Frenkel, {\it Phys. Lett.} {\bf B246} (1990) 75.

\item{[6]}
   V.G. Knizhnik and A.B. Zamolodchikov, {\it Nucl. Phys.}
   {\bf B247} (1984) 83.

\item{[7]}
   P. Christe and R. Flume, {\it Nucl. Phys.} {\bf B282}
   (1987) 466;\par
   P. Christe, PhD Thesis, Bonn (1986).

\item{[8]}
   E. Date, M. Jimbo, A. Matsuo and T. Miwa, in: Yang-Baxter
   Equations, Conformal Invariance and Integrability in
   Statistical Mechanics and Field Theory, World Scientific
   (1989).

\item{[9]}
   V.S. Dotsenko and V.A. Fateev, {\it Nucl. Phys.} {\bf B240}
   (1984) 312; {\bf B251} (1985) 691.

\item{[10]}
   A. Belavin, A. Polyakov and A. Zamolodchikov, {\it Nucl.
   Phys.} {\bf B241} (1984) 333.

\item{[11]}
   F.G. Malikov, B.L. Feigin and D.B. Fuks, {\it Funct. Anal.
   Prilozhen.} 20, no. 2 (1987) 25.

\item{[12]}
   A.B. Zamolodchikov and V.A. Fateev,
   {\it Sov. J. Nucl. Phys.} {\bf 43} (1986) 657.

\item{[13]}
   K. Aomoto, {\it J. Math. Soc. Japan} {\bf 39} (1987) 191.

\item{[14]}
   Vl.S. Dotsenko,    {\it Nucl. Phys} {\bf B358} (1991) 67.

\item{[15]}
   V. G. Kac and D. A. Kazhdan, {\it Adv. Math.} {\bf 34} (1979) 97.

\item{[16]} M. Bauer and N. Sochen,
  {\it Phys. Lett.} {\bf B275}  (1992) 82.

\item{[17]}
   V.K. Dobrev, {\it Lett. Math. Phys.} {\bf 22} (1991) 251.

\item{[18]}
   S. Mukhi and  S. Panda, {\it Nucl. Phys.} {\bf B338} (1990)
   263.

\item{[19]} L. Feh\'er, L. O'Raifeartaigh, P. Ruelle, I.
   Tsutsui and A. Wipf, Dublin preprint, DIAS-STP-91-29.

\item{[20]}
    A. Kent, Cambridge preprint DAMTP-91/31.

\bye